# $CuSbSe_2$ photovoltaic devices with 3% efficiency

Adam Welch[1,2*], Lauryn Baranowski[1,2], Pawel Zawadzki[1], Stephan Lany[1], Colin Wolden[2], Andriy Zakutayev[1*]

[1] National Renewable Energy Laboratory, Golden, CO, USA

[2] Colorado School of Mines, Golden, CO, USA

[*]corresponding authors: adam.welch@nrel.gov, andriy.zakutayev@nrel.gov

Recent technical and commercial successes of existing thin film solar cell technologies motivates exploration of next-generation photovoltaic (PV) absorber materials. Of particular scientific interest are compounds like $CuSbSe_2$, which do not have the conventional tetrahedral semiconductor bonding. Here, we demonstrate 1.5 μm thick $CuSbSe_2$ PV prototypes prepared at 380-410°C by a self-regulated sputtering process using the conventional substrate device architecture. The p-type $CuSbSe_2$ absorber has a 1.1 eV optical absorption onset, ~$10^5$ $cm^{-1}$ absorption coefficient at 0.3 eV above the onset, and a hole concentration of ~$10^{17}$ $cm^{-3}$. The promising >3% energy conversion efficiency ($J_{sc}$ = 20 mA/$cm^2$, FF = 0.44, $V_{oc}$ = 0.35 V) in these initial devices is limited by bulk recombination that limits photocurrent, device engineering issues that affect fill factor, and a photovoltage deficit that likely results from the non-ideal $CuSbSe_2$/CdS band offset.

Current thin film solar cell technologies such as CdTe and Cu(In,Ga)$Se_2$ have achieved champion efficiencies in excess of 20% after decades of research and development, and are being commercialized at a large scale. As such, research and development are free to explore more novel photovoltaic (PV) absorber materials, like $Cu_2ZnSnSe_4$[1] or $Cu_2SnS_3$,[2] both tetrahedrally bonded solids. Such studies of novel inorganic semiconductors can create a more secure energy source by increasing the diversity of the solar energy conversion technologies, with no single chemical element being critical to power generation. In addition, studying non-tetrahedrally bonded PV absorber materials, such as SnS[3] and $Sb_2Ch_3$ (Ch=S,Se)[4,5], can reveal which physical properties, chemical characteristics or structural features are most critical for efficient solar energy conversion. For all these reasons, we choose investigate $CuSbSe_2$ as a PV absorber.

$CuSbSe_2$ is an interesting novel PV absorber material because it is chemically quite similar to the well-known $CuInSe_2$, but structurally very different due to the low-valent state of antimony.[6,7] In $CuSbSe_2$, antimony is in the +III oxidation state like indium in $CuInSe_2$; however, unlike indium, antimony contains two lone pair non-bonding electrons that frustrate the tetrahedral bonding, resulting in a layered orthorhombic (space group Pmnb) chalcostybite crystal structure.[8,9] There has been surprisingly little attention to PV applications of $CuSbSe_2$, compared to $CuSbS_2$.[10,11] The few published studies describe electronic structure (Cu-Se valence band / Sb-S conduction band and $\alpha = 10^4$-$10^5$ $cm^{-1}$)[12,13], as well as the optoelectronic properties of thin films ($E_g$=1.1-1.2 eV, p-type) and photoelectrochemical device performance (~10% EQE, ~50 $\mu A/cm^2$ photocurrent)[14,15] of Cu-rich and Cu-poor $CuSbSe_2$ absorbers.

In this letter, we demonstrate initial prototypes of $CuSbSe_2$ PV devices with a substrate architecture comprised of a CdS front contact and Mo back contact. As previously demonstrated with $CuSbS_2$,[1617] stoichiometric phase-pure $CuSbSe_2$ can be grown using self-regulated, three-stage approach, where the growth rate is controlled by $Cu_2Se$ flux, while excess $Sb_2Se_3$ remains in the vapor phase. The resulting micron scale grains grown between 380°C and 410°C have a 1.1 eV optical absorption onset, a hole density of $10^{17}$ $cm^{-3}$, ~50% external quantum efficiency, and >3% energy conversion efficiency in $CuSbSe_2$. These results are encouraging for a PV device prototype, and call for further research and development of $CuSbSe_2$ PV technology.

The conditions for absorber synthesis were developed through combinatorial sputtering (see Supplementary information for more details). As shown in Fig. 1, two $Sb_2Se_3$ binary targets were oriented perpendicular to a third $Cu_2Se$ source. The glass substrates were intentionally not rotated, resulting in orthogonal gradients of absorber thickness and deposition rate (Fig. 1), since excess $Sb_2Se_3$ remained in the vapor phase. Normalized to deposition rate, substrate regions closer to the $Cu_2Se$ target experienced a lower $Sb_2Se_3$ flux than the regions farther from the $Cu_2Se$ target. This relative $Sb_2Se_3$ flux, $\Delta F$, was quantified as $\Delta F(Sb_2Se_3) = cP(Sb_2Se_3) / (P(Cu_2Se) + \gamma d_t)$, where P are the gun powers, $d_t$ is the throw distance from the $Cu_2Se$ target, $\gamma$ is a fitting parameter that describes reduction in flux density with increasing throw distance, and c is a constant that converts sputtering power to atom flux (assumed here equal for $Cu_2Se$ and $Sb_2Se_3$). The resulting $CuSbSe_2$ absorbers had dense morphology and relatively large (>1 μm) but irregular grains (inset, Fig. 2b).

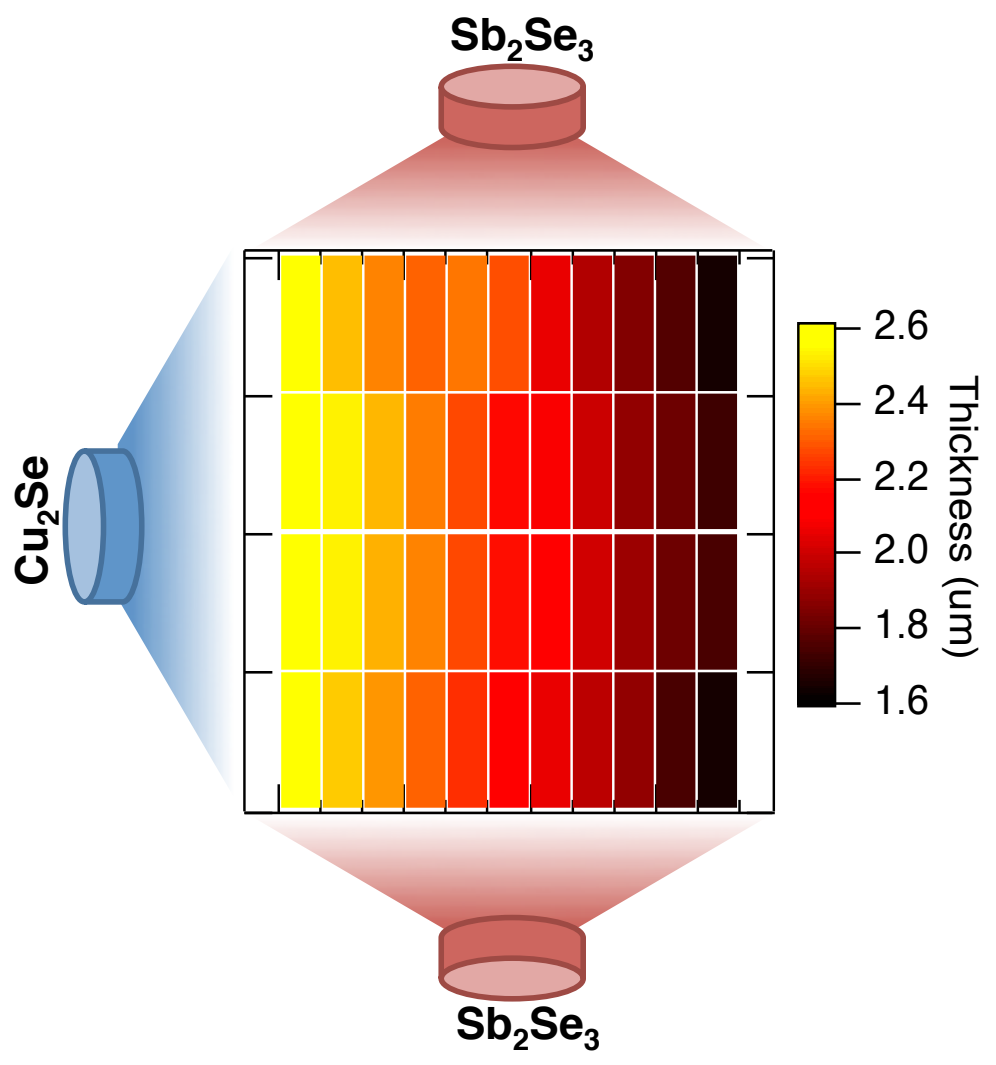

Figure 1: Schematic of the experimental setup showing relationship of sputter sources to substrate, and the resulting XRF measured thickness profile.

The $CuSbSe_2$ material- and device- combinatorial sample libraries fabricated were characterized using spatially-resolved measurement techniques for composition, structure, thickness and device performance. A total of 44 spatially-resolved automated measurements were taken on each combinatorial sample library, followed by single-point manual characterization of morphology, optical and electrical properties on the most interesting samples. For the absorber materials libraries, 4 equivalent rows of 11 data points were measured, each with different thicknesses, as shown in Fig. 1. For the PV device libraries, 11 equivalent data points in 4 groups for devices were measured, orthogonal to what is shown in Fig. 1. More fabrication and characterization details for the $CuSbSe_2$ materials- and device- combinatorial libraries are summarized in Supplementary Information, and were similar to our recent $CuSbS_2$,[16,17] and $Cu_2SnS_3$[18,19] publications.

Ab-initio calculations were performed using the VASP code,[20],. The band gap and optical absorption spectrum was theoretically determined from quasiparticle energy calculations in the GW approximation.[21] In order to determine the single-phase stability regions of $CuSbSe_2$, we performed density functional theory (DFT) calculations for $CuSbSe_2$ and the relevant competing phases, including $Sb_2Se_3$, $Cu_3SbSe_3$, and $Cu_2Se$. The respective compound formation enthalpies were determined using the fitted elemental reference energies.[22] The vapor pressures of $Sb_2Se_3$(g) in equilibrium with $CuSbSe_2$(s) were calculated for the $3CuSbSe_2$(s) -> $Cu_3SbSe_3$(s) + $Sb_2Se_3$(g) and $2CuSbSe_2$(s) -> $Cu_2Se$ (s) + $Sb_2Se_3$(g) decomposition pathways. To calculate the free energy change for these reactions we used DFT formation enthalpies of solid phases and the free energy of $Sb_2Se_3$(s) sublimation calculated from experimental vapor pressure.[23]

Previous experimental work by our group[16,17] has demonstrated phase pure growth of the related $CuSbS_2$ compound, by employing substrate temperatures above the sublimation of $Sb_2S_3$ but below the decomposition into Cu-rich phases. The same synthesis strategy has been applied in this work for $CuSbSe_2$ absorber. As shown in theoretical analysis in Fig. 2a, the shaded yellow region between $CuSbSe_2$ decomposition and $Sb_2Se_3$ sublimation defines the processing window for phase-pure material. At a given temperature, there is a range of $Sb_2Se_3$ equilibrium vapor pressure between the sublimation and decomposition lines for the $CuSbSe_2$ phase. If the $Sb_2Se_3$ partial pressure is too high, $Sb_2Se_3$ impurities will remain in the film. If the $Sb_2Se_3$ partial pressure is too low, the deposited $CuSbSe_2$ will decompose to $Cu_3SbSe_3$. Further reduction in $Sb_2Se_3$ partial pressure produces no ternary phases. Thus, phase-pure $CuSbSe_2$ should be grown in the temperature-pressure region bound by these two decomposition lines (Fig. 2a).

Experimentally, the $Sb_2Se_3$ pressure is controlled in our chamber by the relative $Sb_2Se_3$ flux (see definition above). Consistent with the theoretical calculations (Fig. 2a), we find that for a given substrate temperature (350C, 380C and 410C in our experiments), there is a range of relative $Sb_2Se_3$ flux values which produce stoichiometric $CuSbSe_2$ material with a Cu atomic ratio of Cu/(Cu+Sb+Se)=0.25 (Fig. 2b) and phase-pure XRD pattern (Fig. 3a). The too low relative $Sb_2Se_3$ flux leads to $CuSbSe_2$ decomposition into $Cu_2Se$ + $Cu_3SbSe_3$, resulting in Cu atomic ratios >0.25, and too high $Sb_2Se_3$ flux leads to $Sb_2Se_3$ impurities remain in the film, resulting in Cu atomic ratios < 0.25. Projecting experimental Fig. 2b onto theoretical Fig. 2a, it appears that the experimental $Sb_2Se_3$ relative flux ratios achieved in our chamber roughly translate to an $Sb_2Se_3$ partial pressure range of $10^{-8}$ – $10^{-5}$ Torr, depending on the $Cu_2Se$ throw distance. This graded $Sb_2Se_3$ flux ratio along the combinatorial library results in thicker films and higher Cu chemical potential closer to the $Cu_2Se$ target. Combinatorial studies to decouple these two gradients and to study their effects on $CuSbSe_2$ device performance are currently in progress and will be reported in a separate full-length article.

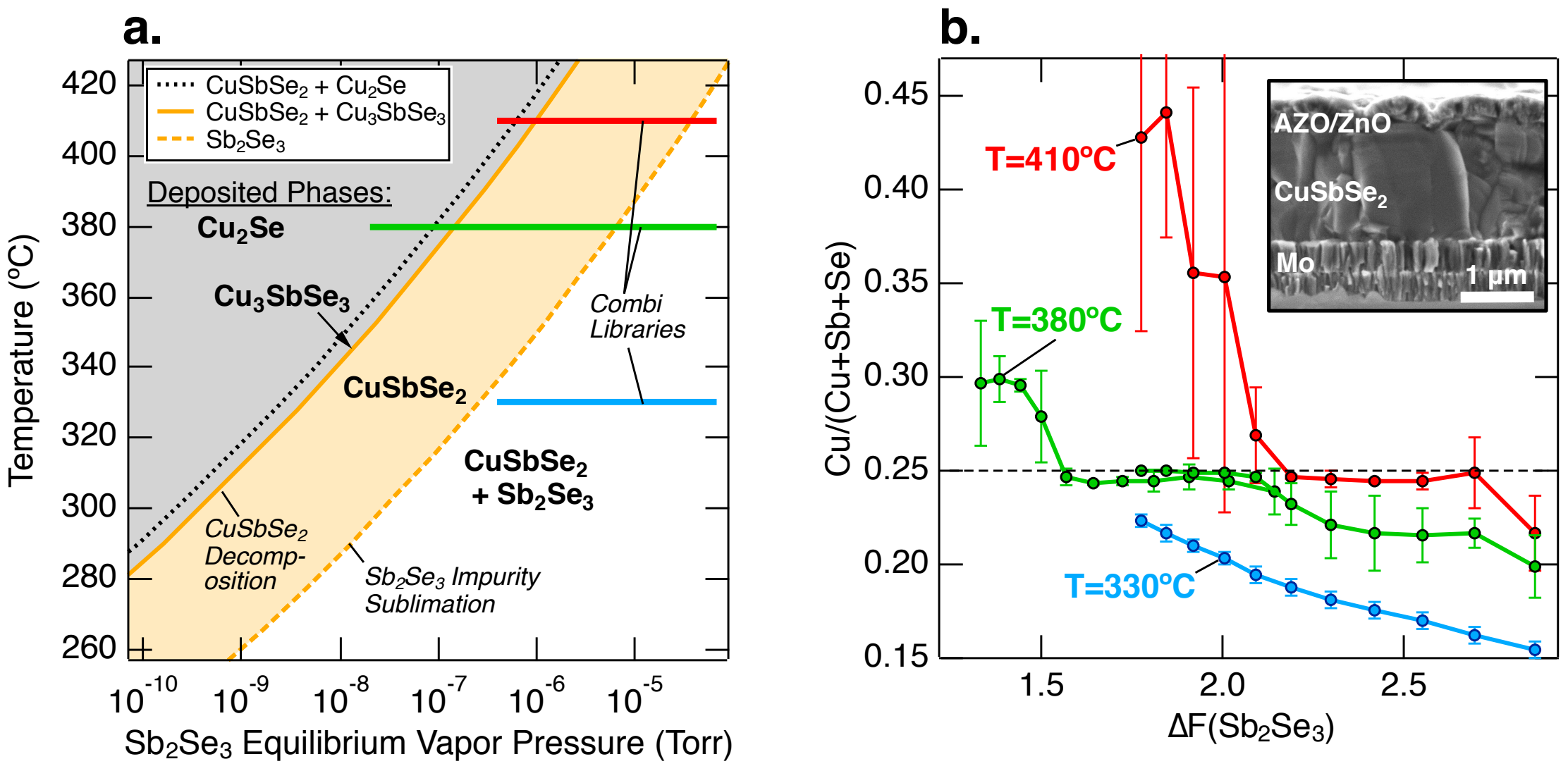

Figure 2: (a) Theoretical calculations reveal a range of $Sb_2Se_3$ equilibrium vapor pressure and temperature where phase-pure $CuSbSe_2$ can be grown. (b) Experiments confirm that stoichiometric $CuSbSe_2$ can be deposited in a range of $Sb_2Se_3$ fluxes and substrate temperatures. Inset: cross-section SEM image, displaying film thickness, device structure and absorber morphology.

Regions of the combinatorial libraries corresponding to the different amounts of Cu incorporation were characterized for optical absorption to identify the band gap of the phase pure material, and to study the effect of impurity phases on the optical properties. As shown in Fig. 3b, the stoichiometric $CuSbSe_2$ films had a sharp absorption onset at 1.1 eV that saturated to nearly $10^5$ $cm^{-1}$ only 0.35 eV above the band gap. These results are in very good agreement with the GW theoretical calculations, leading to high confidence in the reported $CuSbSe_2$ optical absorption properties. This is indicative of a nearly-direct band gap in this material, consistent with the prior theoretical[12,13] and experimental[14,15] reports. $Cu_2Se$ impurities appear to increase sub-bandgap absorption in $CuSbSe_2$, consistent with additional carriers from this degenerate semiconductor. Conversely, $Sb_2Se_3$ impurities seem to further reduce sub-band absorption in $CuSbSe_2$, possibly due to total elimination of $Cu_2Se$ impurities. However it should be noted that the $10^3$ $cm^{-1}$ values are close to the instrument error for the measured very thin ~500nm films, so thicker films would be required to study the sub-gap absorption in more details in the future.

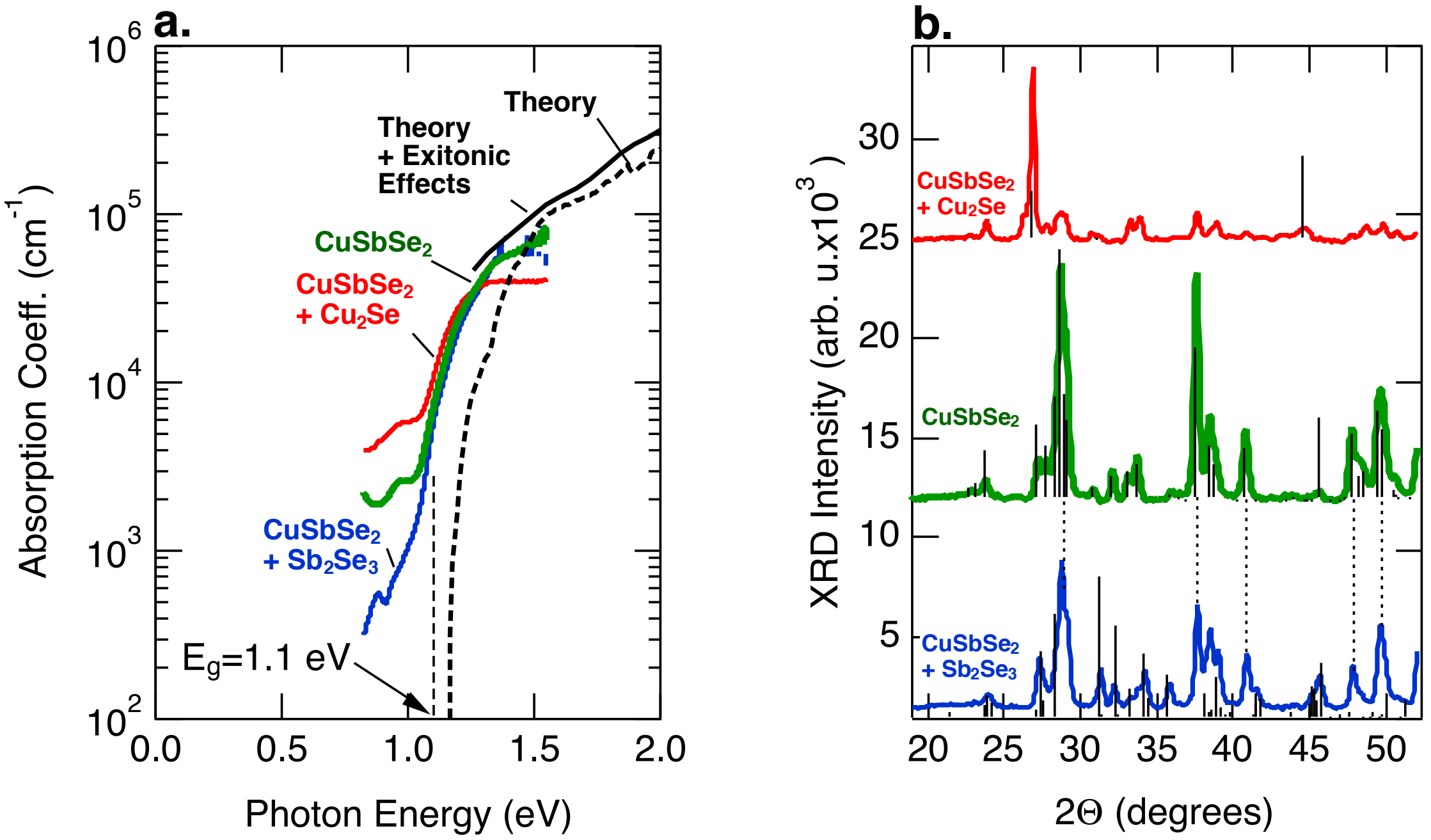

Figure 3: (a) Absorption coefficient values for stoichiometric $CuSbSe_2$, and for $CuSbSe_2$ that contains phase impurities. Black lines are the results of GW theoretical calculations. (b) XRD phase identification for the different regions on the combinatorial libraries. Black lines are the reference patterns for $CuSbSe_2$ (green), $Sb_2Se_3$ (blue), and $Cu_2Se$ (red).

Figure 4a displays the JV characteristic of the highest efficiency $CuSbSe_2$ PV device, along with a histogram analysis of all the nominally uniform, not shunted devices from the thickest (1.5 μm) row of a combinatorial library grown at 380 °C. Excluding the 2 shunted devices, the 9 remaining devices had average values of: $V_{oc}$ = 346±7mV, $J_{sc}$ = 20.5±1.7 mA/cm$^2$, FF = 43.9±2.8, and efficiency = 3.12±0.41%. The difference between the $J_{sc}$ measured by J-V (up to 22 mA/cm$^2$, Fig. 4a) and integrated EQE x AM1.5G (up to 18 mA/cm$^2$, Fig. 4b) is likely due to asymmetry in the $CuSbSe_2$ spectral response combined with 400-500 nm peaks in the Xe-lamp spectrum. The thinner (1.3 μm) row of devices had slightly lower efficiencies due to lower photocurrents. The remaining two rows of devices were used for other experiments, and hence are not discussed here.

The $J_{sc}$, $V_{oc}$ and FF values listed above are approximately a half of what might be expected for an ideal 1.1 eV gap absorber. This is quite promising as for an initial $CuSbSe_2$ PV device prototype, but calls for further analysis to improve this performance. The less-than-ideal FF is probably due to the lack of device optimization, leading to shunts, isolation issues and so on, which are quite typical of the initial PV device prototypes. The measured non-ideal $V_{oc}$ can be tentatively attributed in part to the extrinsic device problem, namely the likely cliff-type band offset between the n-type CdS heterojunction and the layered ternary $CuSbSe_2$ absorber (similar to $CuSbS_2$[17]). This is in contrast to tetrahedrally-bonded quaternary $Cu_2ZnSnS_4$[24,25] or $Cu_2SnS_3$,[26,27] where similar $V_{oc}$ deficit is often attributed to the intrinsic materials problem of cation disorder. Finally, the half-of-ideal short circuit currents ($J_{sc}$) indicate substantial room for improvement in collection of photogenerated charge carriers, given that the 1.5 μm thick $CuSbSe_2$ layers are not absorption limited (Fig. 3b)

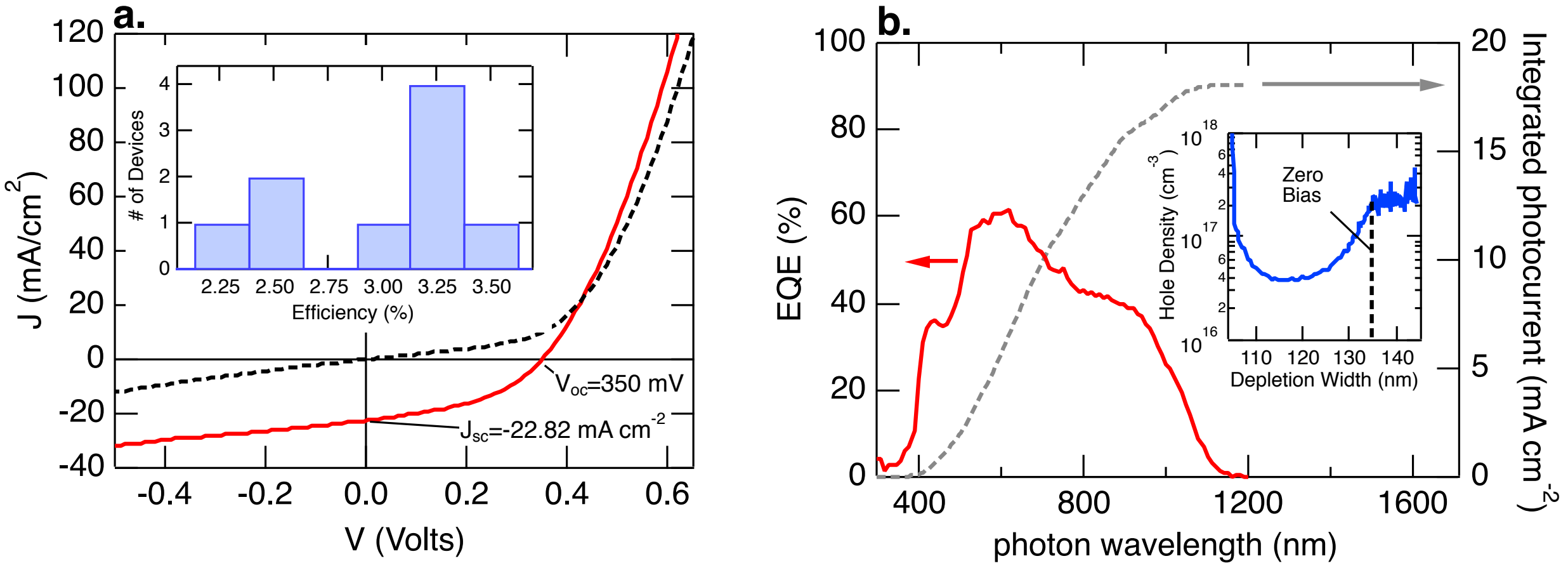

Figure 4: (a) JV results for the most efficient 3.5% $CuSbSe_2$ PV device and a histogram analysis of efficiency for all 9 nominally equivalent, non-shunted devices from the first row on the combinatorial library as the inset. (b) EQE results for the 3.5% $CuSbSe_2$ PV device, including the integrated EQE x AM1.5G product, and CV analysis as the inset.

To distinguish between the bulk- and interface photocurrent recombination pathways in the 3.5% $CuSbSe_2$ PV devices, we performed spectrally-resolved EQE measurement under white light bias (Fig. 4b). As expected, the EQE for > 1000nm and <500 nm is limited by insufficient absorption in $CuSbSe_2$ and parasitic absorption in CdS, respectively. In the 500 - 1000 nm region of interest, the EQE results indicate 40% collection between 750 and 1000 nm wavelength, suggestive of significant recombination losses in the bulk. The collection improves up to 60% in the 500 - 750 nm spectral range, pointing to drift-enhanced carrier collection in the space charge region. Consistent with this result, the CV characteristics indicate a narrow depletion width of 135 nm due to relatively high hole concentration of $2x10^{17}$ $cm^{-3}$. Given the measured absorption coefficient ($7x10^4$ $cm^{-1}$ at 900nm, Fig. 3a), a Beer-Lambert absorption depth calculation suggest that 50% of the <900nm photons are absorbed within the 135 nm space charge regions. This 50% number is comparable with the average EQE values in the 500-1000 nm spectral range, supporting the hypothesis of drift-enhanced operation of our $CuSbSe_2$ PV device.

In summary, phase-pure $CuSbSe_2$ thin films have been synthesized by sputtering, characterized for optical absorption, and made into initial PV device prototypes. The stoichiometric $CuSbSe_2$ is relatively easy to grow using the self-regulated synthesis technique that keeps excess $Sb_2Se_3$ in the vapor phase. The $CuSbSe_2$ absorption onset is indicative of nearly-direct 1.1 eV band gap that matches well the solar spectrum, making this material interesting for thin film PV device integration. Preliminary PV devices in the substrate architecture with Mo back contact and CdS heterojunction partner show promising >3% initial energy conversion efficiencies, limited by bulk photocurrent recombination (affects $J_{sc}$), likely cliff-type $CuSbSe_2$/CdS band offset (affects $V_{oc}$), and device engineering issues (affects FF). Therefore, future work should focus on identification and improvement of band offsets between CdS and $CuSbSe_2$ to increase $V_{oc}$, decreasing bulk defect recombination of the $CuSbSe_2$ absorber material to increase $J_{sc}$, and usual device engineering improvements to increase the FF.

The “Rapid Development of Earth-abundant Thin Film Solar Cells” project is supported by the U. S. Department of Energy, Office of Energy Efficiency and Renewable Energy, as a part of the SunShot initiative, under Contract No. DE-AC36-08GO28308 to NREL. We would like to acknowledge Clay DeHart for consistent high quality and rapid application of top contact layers, and F. Willian de S. Lucas for informative discussion.

**Supplementary information**

*Absorber synthesis details*. For $CuSbSe_2$ absorber synthesis, we leverage our recently established combinatorial approach to absorber materials research and PV device development, initially demonstrated on the example of the related $CuSbS_2$ materials. Depositions were performed in a sputtering chamber with $3x10^{-3}$ Torr of Ar and $10^{-7}$ Torr base pressure. The films were grown from 2 $Sb_2Se_3$ and 1 $Cu_2Se$ targets, each 50 mm in diameter. The absorbers were deposited on 50x50 mm soda lime glass (SLG), or SLG/Mo substrates held at 350-410C during the deposition, and cooled/heated in excess $Sb_2Se_3$ flux.

*Device fabrication details*: The $CuSbSe_2$ PV devices were fabricated by DC sputter deposition of a bilayer Mo on SLG substrates, followed by sputter-deposition of the $CuSbSe_2$ absorber deposition as described above. The front contact was formed by chemical bath deposition of CdS buffer layer, with subsequent RF sputter deposition of the i-ZnO/ZnO stack and e-beam evaporation of Ni:Al metal fingers and MgF antireflection coating. The devices from contacts were isolated from each other by razor blade scribing.

*Characterization details*. The resulting combinatorial libraries of the absorber material and PV devices were studied by spatially-resolved characterization techniques, including AM 1.5G J-V measurements of the device performance (calibrated by Si reference cell), X-ray diffraction (XRD) for absorber crystallographic structure and phase composition, x-ray fluorescence (XRF) for absorber chemical composition and thickness (as verified by Dektak profilometer). For several selected points on the libraries we also measured optical absorption properties, cross-section scanning electron microscopy (SEM), external quantum efficiency (EQE), and capacitance-voltage (CV) analysis.  Data processing for all tools was done with custom routines implemented in IgorPro software package.

*Computational details*. The ab initio calculations in this work were performed with the Vienna Ab initio Simulation Package (VASP) using the the projector-augmented wave (PAW) implementation of density functional theory (DFT)[Error! Bookmark not defined.] and many-body perturbation theory in the GW approximation[21] The Perdew-Burke-Ernzerhof (PBE) generalized gradient approximation (GGA) was used for the DFT exchange-correlation functional[28]. The GGA+U method[29] with U = 5 eV was employed to for Cu-d orbitals. The detailed computational settings used for the GW band gap calculations are identical to those published before, including the use of an onsite-potential to account for the underbinding of Cu-d states in GW.[30] The optical spectra including excitonic effects were calculation within time-dependent DFT, using a hybrid functional kernel[31] with a fraction of Fock exchange of $\alpha = 1/\varepsilon$, where $\varepsilon = 13$ is the static electronic dielectric constant obtained from the GW calculation.

[25] Schorr, Susan, and Gabriela Gonzalez−Aviles. "In−situ investigation of the structural phase transition in kesterite." *physica status solidi (a)* 206, no. 5 (2009): 1054-1058.

[26] Zawadzki, Paweł, Andriy Zakutayev, and Stephan Lany. "Entropy-Driven Clustering in Tetrahedrally Bonded Multinary Materials." *Physical Review Applied* 3, no. 3 (2015): 034007.

[27] Baranowski, Lauryn L., Kevin McLaughlin, Pawel Zawadzki, Stephan Lany, Andrew Norman, Hannes Hempel, Rainer Eichberger, Thomas Unold, Eric S. Toberer, and Andriy Zakutayev. "Effects of disorder on carrier transport in $Cu_2SnS_3$." *arXiv preprint arXiv:1504.01327* (2015).

[28] Perdew, John P., Kieron Burke, and Matthias Ernzerhof. "Generalized gradient approximation made simple." *Physical review letters* 77.18 (1996): 3865.

[29] Dudarev, S. L., et al. "Electron-energy-loss spectra and the structural stability of nickel oxide: An LSDA+ U study." *Physical Review B* 57.3 (1998): 1505.

[30] Lany, Stephan. "Band-structure calculations for the 3 d transition metal oxides in G W." *Physical Review B* 87.8 (2013): 085112.

[31] Paier, Joachim, Martijn Marsman, and Georg Kresse. "Dielectric properties and excitons for extended systems from hybrid functionals." *Physical Review B*78.12 (2008): 121201.